\documentclass[reprint,aps,preprintnumbers,amsmath,amssymb,superscriptaddress]{revtex4-2}

\usepackage{graphicx}
\usepackage{dcolumn}
\usepackage{bm}
\usepackage{multirow}
\usepackage{xcolor}
\usepackage[normalem]{ulem}
\usepackage{amsmath}
\usepackage{gensymb}
\usepackage[colorlinks=true]{hyperref}
\usepackage{booktabs}
\usepackage{threeparttable}
\usepackage{mathptmx}

\newcommand{\changes}[1]{{\color{black} #1}}

\begin{document}


\title{Role of a Quarter-Wave Plate in Confocal Microscopy: Signature of Spin-Orbit Interactions}


\author{Wenze Lan}
\affiliation{Institute for Condensed Matter Physics, Technische Universität Darmstadt, Hochschulstr. 6, 64289 Darmstadt, Germany}
\author{Anton L\"ogl}
\affiliation{Institute for Condensed Matter Physics, Technische Universität Darmstadt, Hochschulstr. 6, 64289 Darmstadt, Germany}
\author{Meryem Benelajla}
\affiliation{eyeo BV, High Tech Campus 41, 5656 AE Eindhoven, The Netherlands}
\author{Clemens Sch\"afermeier}
\affiliation{attocube systems GmbH,  Eglfinger Weg 2, 85540 Haar, Germany}
\author{Khaled Karrai}
\affiliation{attocube systems GmbH,  Eglfinger Weg 2, 85540 Haar, Germany}
\author{Bernhard Urbaszek}
\email{bernhard.urbaszek@pkm.tu-darmstadt.de}
\affiliation{Institute for Condensed Matter Physics, Technische Universität Darmstadt, Hochschulstr. 6, 64289 Darmstadt, Germany}



\date{\today}




\begin{abstract}
\changes{Spin–orbit interactions of light couple polarization and spatial degrees of freedom, underpinning phenomena such as the spin Hall effect of light. Although widely explored at interfaces and in tightly focused beams, their impact in nominally paraxial confocal systems remains largely unexamined. Here we show that a single quarter-wave plate embedded in a simple confocal geometry between polarizers can strongly reshape the transverse structure of a Gaussian beam. We observe an enhancement of the polarization extinction ratio by more than two orders of magnitude, accompanied by a transformation of the Gaussian intensity profile into a first-order Hermite–Gaussian-like two-lobe mode. The orientation of this pattern is continuously tunable via rotation of the wave plate, evidencing polarization-controlled reorientation of the transverse field. To explain these observations, we introduce a minimal extension of Jones matrix formalism incorporating complex parameters that quantitatively reproduces the measurements. Our results uncover a previously overlooked form of spin–orbit-mediated mode control in standard confocal optics and establish a simple route to on-demand spatial mode engineering for applications in resonant spectroscopy, optical imaging and quantum optics.}

\end{abstract}


\maketitle

\textbf{Introduction.} Spin-orbit interactions (SOI) of light constitute a general framework describing the coupling between polarization and spatial degrees of freedom \cite{bliokh2015spin,shen2019optical}. These interactions originate from the inherently vectorial nature of electromagnetic waves and become particularly relevant when polarization-dependent phase or amplitude variations are linked to the spatial structure of a beam. 
Optical SOI underlie a broad range of phenomena, including geometric (Pancharatnam-Berry) phase effects \cite{berry1984quantal,vinitskij1990topological,bhandari1997polarization}, spin-dependent beam shifts \cite{bliokh2008geometrodynamics,hosten2008observation}, and spin-to-vortex conversion \cite{zhao2007spin,Bliokh:11}. 
Beyond these classical effects, SOI underpins emerging topics including the formation of topological polarization singularities \cite{liu2019circularly,zeng2021dynamics,qin2023arbitrarily} and the generation of optical skyrmions \cite{shen2024optical,hakobyan2025q}. Understanding and harnessing SOI is therefore crucial for advanced applications in nanophotonics, optical manipulation, and quantum information processing.

A paradigmatic manifestation of optical SOI is the spin Hall effect of light (SHEL) \cite{onoda2004hall,bliokh2013goos,toppel2013goos,ling2017recent,kim2023spin}, in which left- and right-handed circularly polarized components experience opposite transverse shifts upon reflection or transmission. This effect was originally investigated at planar isotropic interfaces \cite{bliokh2006conservation,hosten2008observation}, where it is closely related to the transverse Imbert-Fedorov shift \cite{imbert1972calculation} and to longitudinal beam shifts associated with the Goos-Hänchen effect \cite{goos1947neuer}. Subsequent studies have shown that analogous spin-dependent effects can also arise in a broader class of systems exhibiting polarization-dependent optical responses, including anisotropic or polarization-selective elements such as uniaxial crystal plates \cite{bliokh2016spin}, polymers \cite{takayama2018enhanced}, linear polarizers \cite{korger2014observation,bliokh2019spin}, and metamaterials \cite{takayama2018photonic}. In these systems, the conventional Fresnel coefficients are generalized to polarization-dependent transmission or reflection matrices, highlighting that optical SOI is not restricted to simple interfaces but can emerge more broadly in systems exhibiting polarization-dependent optical responses. \changes{Among such systems, the quarter-wave plate (QWP) provides a particularly simple and widely used example, as its birefringent phase retardation introduces a phase difference between orthogonal polarization components, thereby enabling SOI effects.}

Experimentally, the SHEL has most commonly been accessed and amplified using focused beams and quantum weak measurement techniques \cite{ritchie1991realization,hosten2008observation,dressel2014colloquium,choi2024experimental,lee2025real}, where small spin-dependent momentum-space shifts are converted into measurable spatial displacements upon propagation. By contrast, in confocal optical systems ubiquitous in microscopy \cite{rodriguez2010optical,steindl2023cross}, spectroscopy \cite{kuhlmann2013dark,shree2021guide}, and quantum optics \cite{nick2009spin}, the excitation and detection beams are typically well collimated over most of the optical path and traverse a sequence of polarization optics before being focused onto the sample and collected through a pinhole in a conjugate focal plane. In this widely used configuration, Benelajla et al.~\cite{benelajla2021physical} reported that a confocal arrangement incorporating a reflecting surface between the polarizer and analyzer can achieve a cross-polarization extinction ratio above $10^8$, exceeding the intrinsic polarizer extinction by over three orders of magnitude, as a consequence of the SOI of light. Building on this concept, Steindl et al. \cite{steindl2023cross} subsequently reported a tenfold improvement in single-photon contrast in cryogenic confocal microscopy. \changes{However, these investigations focused primarily on the role of reflecting surfaces, leaving the role of transmissive polarization optics largely unexplored. Despite its intrinsic anisotropy,} a QWP is conventionally regarded as a purely polarization-compensating element \cite{kuhlmann2013dark}, introduced to correct residual ellipticity and enhance the linear polarization extinction, without affecting the spatial mode structure of the beam. Whether and how optical SOI can manifest in such confocal geometries incorporating a QWP has, however, remained largely unexplored, despite previous studies on spin-dependent effects induced by a \textit{tilted} half-wave plate in non-confocal configurations \cite{bliokh2016spin}.

\begin{figure*}[htbp]
	\centering
	\includegraphics[width=1.0\textwidth]{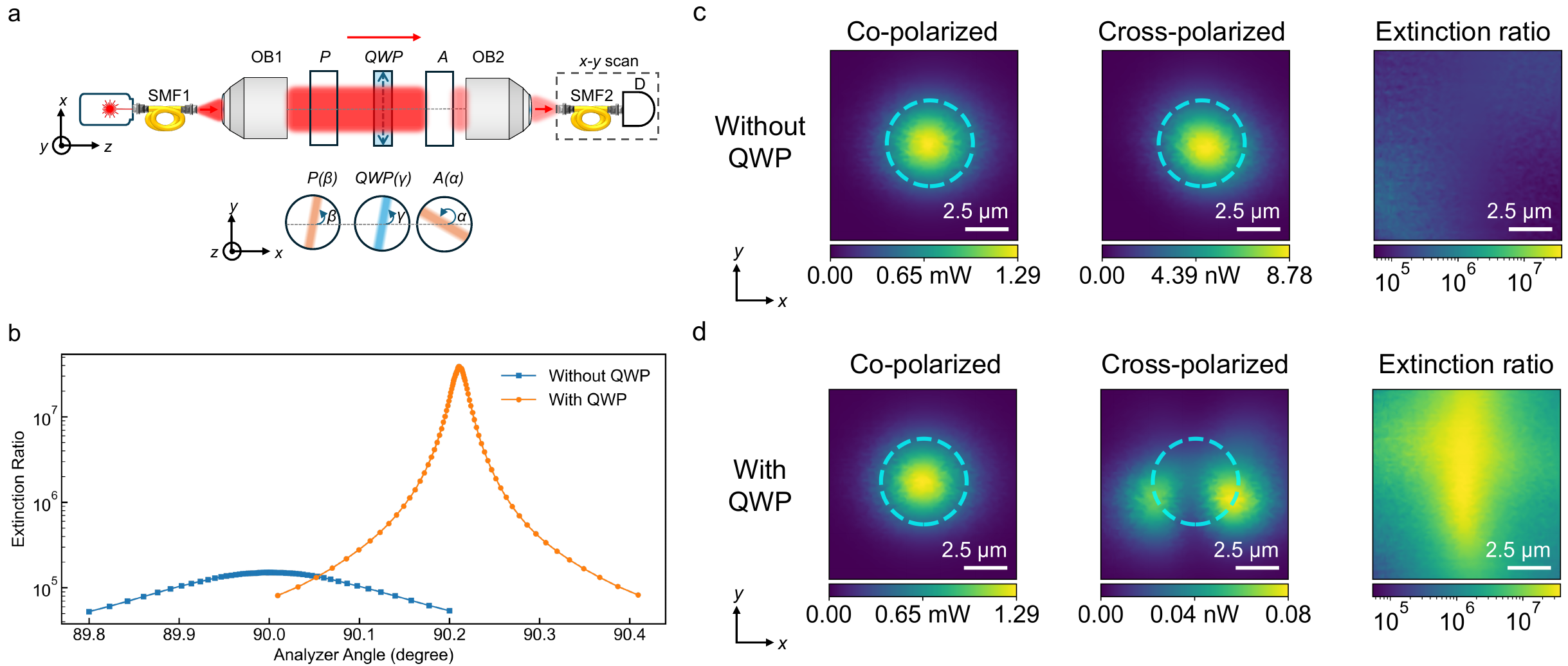}
	\caption{\changes{
\textbf{Experimental configuration and polarization-dependent confocal response.}
\textbf{a} Schematic of the confocal setup. A Gaussian beam is prepared by a polarizer ($P$) at angle $\beta$ and transmitted through a quarter-wave plate (QWP) with fast-axis angle $\gamma$, followed by an analyzer ($A$) at angle $\alpha$. The transmitted field is coupled into a single-mode fiber (SMF2) and detected by a photodiode, while the fiber position is scanned in the focal plane. Unless otherwise stated, all elements operate at normal incidence.
\textbf{b} Extinction ratio curves are shown for both cases, with (orange) and without (blue) a QWP. \changes{The polarizer, QWP and analyzer are initially set to $\beta = 0^\circ$, $\gamma = 0^\circ$ and $\alpha=90^\circ$, respectively.} After optimization, the extinction ratio is enhanced to $\sim 10^7$, compared to $\sim 10^5$ without a QWP. In addition, the extinction maximum is shifted by $\sim 0.209^\circ$ relative to the case without a QWP.
Co- and cross-polarized confocal mappings without (\textbf{c}) and with (\textbf{d}) a QWP are shown. The extinction ratio maps in \textbf{c} and \textbf{d} are obtained by dividing, pixel by pixel, the co-polarized data by the cross-polarized data for both cases. In cross-polarization, mode splitting into two lobes along the $x$ direction is observed when a QWP is introduced. The cyan dotted circle indicates the non-convoluted focal spot 1/e$^2$ waist diameter at the collecting fiber end. }}
\label{fig:1}
\end{figure*}

In this paper, we demonstrate that even a single QWP inserted into a simple confocal geometry can profoundly reshape the transverse structure of a Gaussian beam under cross-polarization conditions. In a minimal optical configuration [see Fig.~\ref{fig:1}a] consisting of a laser source, single-mode fibers, objective lenses, and a ``\textit{polarizer-QWP-analyzer}'' sequence, we observe an enhancement of the polarization extinction ratio by more than two orders of magnitude upon insertion of the QWP, as compared to only using polarizer and analyzer. Spatially resolved measurements at the focal plane reveal that this enhancement is accompanied by a pronounced transformation of the focused field: the initially Gaussian intensity profile is converted into a first-order Hermite-Gaussian (HG)-like two-lobe structure. Remarkably, the resulting two-lobe pattern is not constrained to the transverse coordinate axes, as would be expected for pure HG$_{01}$ or HG$_{10}$ modes. Instead, its orientation can be continuously tuned by rotating the fast axis of the QWP, while the input linear polarization is chosen to be nearly parallel or perpendicular to the QWP fast axis. This behavior is most naturally interpreted as a polarization-dependent reorientation of the transverse field distribution, enabled by the combined action of the QWP and cross-polarization detection. Our results thus uncover a nontrivial regime of SOI in confocal optical systems and challenge the common assumption that a QWP acts solely on the polarization degree of freedom without influencing spatial mode structure. Beyond the fundamental implications, our findings suggest new opportunities for controllable spatial mode shaping and polarization-assisted beam engineering using widely available optical components. We also raise open questions regarding the completeness of standard descriptions of light propagation in polarization-resolved optical systems, particularly when high extinction ratios and spatially structured detection are involved.

\textbf{Enhanced polarization extinction ratio.}
We employ a simplified confocal optical arrangement, schematically shown in Fig.~\ref{fig:1}a, designed to isolate the essential physics underlying the enhanced laser rejection (see Supplementary Methods for details). We first highlight a key experimental observation. When only the polarizer and analyzer are used, the extinction ratio is limited to approximately $10^5$, as indicated by the blue curve in Fig.~\ref{fig:1}b, consistent with the intrinsic polarization leakage of the commercial polarizers. In contrast, inserting a zero-order QWP between the polarizer and analyzer enhances the extinction ratio by more than two orders of magnitude, as shown by the orange curve in Fig.~\ref{fig:1}b.

To account for this behavior, we model the collimated laser beam as a plane wave and analyze the system using the Jones matrix formalism. This framework links small angular deviations in the analyzer to the residual polarization leakage $b$ of the polarizers, as derived in Supplementary Note~\ref{sup_note_1}. Experimentally, we observe an analyzer angle offset of approximately $0.209^\circ$ between the configurations with and without the QWP. From this value, we estimate $b=|\cos(90^\circ+0.209^\circ)| \approx 3.6 \times 10^{-3}$, which is consistent with previously reported values \cite{benelajla2021physical,steindl2023cross}. The corresponding extinction ratio for the configuration without the QWP is therefore $1/b^2 \approx 0.8 \times 10^5$, in good agreement with the experimental result in Fig.~\ref{fig:1}b. In terms of practical leakage suppression performance using a QWP, the maximum extinction ratio in our implementation remains limited to approximately $10^7$, in contrast to the substantially higher values achievable using a reflecting surface \cite{benelajla2021physical}. This limitation arises from the combined effect of a retardance mismatch between the laser wavelength and the QWP design wavelength, together with the finite angular tuning range and discrete step size of the polarizer and analyzer during the optimization process (see Supplementary Fig.~\ref{fig:S1}).

The same Jones matrix analysis can be extended to alternative configurations, such as replacing the QWP with a zero-order half-wave plate (HWP). In this case, no combination of the polarizer and analyzer angles $\alpha$ and $\beta$ can fully suppress the residual leakage (see Supplementary Note~\ref{sup_note_1} and Fig.~\ref{fig:S2}). We further note that Bliokh et al. observed two-lobe HG-like intensity distributions with a central “dark point” under cross-polarization using quantum weak measurement, even when employing a multi-order HWP~\cite{bliokh2016spin}. These observations indicate that a plane-wave Jones matrix description is insufficient to capture the full behavior of the system, as it neglects the transverse spatial structure of the beam. In particular, the enhanced extinction ratio observed in our experiment arises only within the confocal detection geometry, where the transmitted field is both spatially transformed and subsequently filtered by the single-mode fiber. Such geometry-dependent behavior lies beyond a purely polarization-based formalism and points to the essential role of transverse mode evolution. To clarify this mechanism, we analyze in the following section the spatial profile of the transmitted field and its modal transformation under cross-polarization detection.

\begin{figure*}[htbp]
	\centering
	\includegraphics[width=0.85\textwidth]{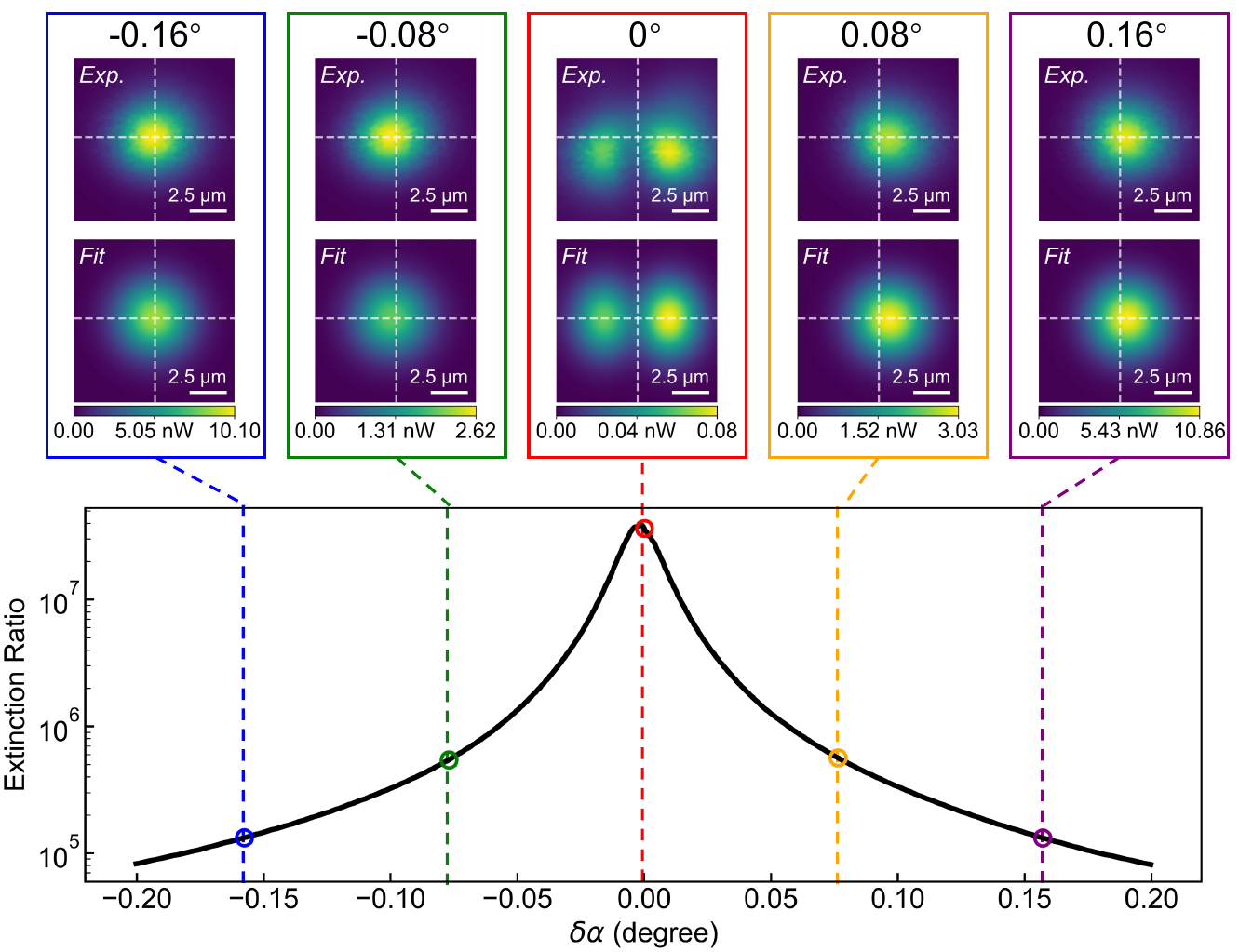}
	\caption{\textbf{Confocal mapping of \changes{linearly polarized (\(\beta = 0^\circ\))} Gaussian laser beams with a QWP for different analyzer angles under cross-polarization.} Measured and fitted evolution of the modal confocal mapping for different analyzer angles $\delta\alpha$. The extinction ratio curve is also presented to guide the extinction levels. At maximum extinction ratio, the mode splits into two lobes at $\delta\alpha=0$. With comparable extinction between $10^5$ and $10^6$ at angles $\delta\alpha=-0.08^\circ$ and $0.08^\circ$, the mode shifts opposite along $x$ axis. When the extinction ratio close to $10^5$ at angles $\delta\alpha=-0.16^\circ$ and $0.16^\circ$, the beam shifts back to center position indicated by two dashed white lines. } 
	\label{fig:2}
\end{figure*}

\begin{figure*}[htbp]
	\centering
	\includegraphics[width=1.0\textwidth]{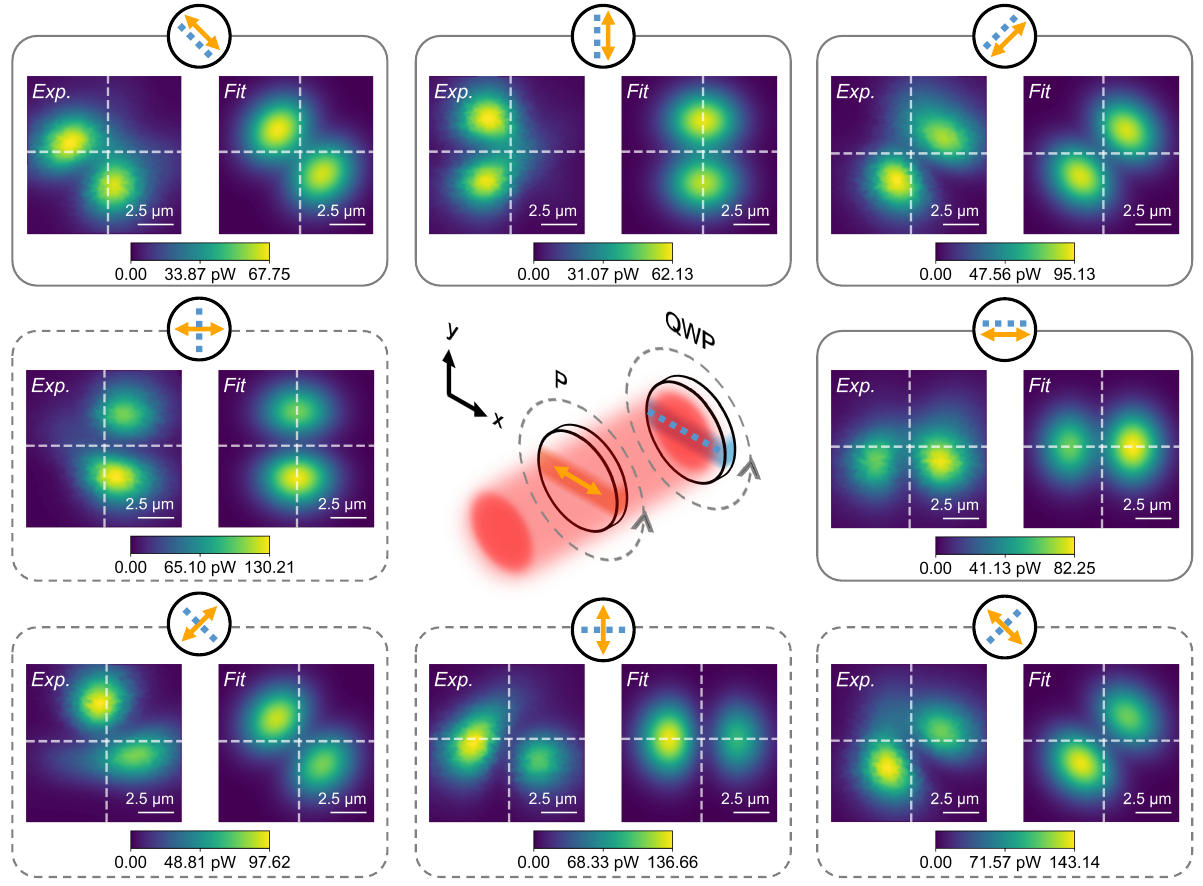}
	\caption{\textbf{Control of first-order HG-like mode orientation.} Measured and fitted cross-polarized confocal modal maps for different combinations of the incident polarization and the QWP fast-axis orientation. \changes{Panels with solid frames correspond to incident linear polarization aligned parallel to the QWP fast axis, while panels with dotted frames correspond to incident linear polarization perpendicular to the QWP fast axis. The center positions of the co-polarization references are indicated by two dashed white lines. The orange arrow and blue dotted line indicate the polarization direction and QWP fast axis, respectively.}}
	\label{fig:3}
\end{figure*}

\textbf{Modal transformation of a transmitted polarized Gaussian beam.}
To identify the origin of the enhanced polarization extinction ratio, we map the detected intensity by scanning the spatial position of the collecting fiber in the focal plane of the objective. \changes{In this measurement, the spatial filtering is set by the Gaussian mode of the collecting single-mode fiber, so that the detected signal is determined by the overlap integral between the focal-plane field and the fiber mode.} In the absence of the QWP between the polarizer and analyzer, the co-polarized and cross-polarized confocal maps in Fig.~\ref{fig:1}c exhibit a pure HG$_{00}$-like Gaussian mode. The corresponding extinction-ratio map in Fig.~\ref{fig:1}c shows an attenuation level of around $10^5$, consistent with the polarizer leakage specification. In contrast, when a QWP is inserted, the co-polarized map in Fig.~\ref{fig:1}d remains Gaussian, whereas the cross-polarized map in Fig.~\ref{fig:1}d reveals a clear splitting of the mode into two lobes. We find an ``intensity hole" at the location of the optical fiber center due to which the light is no longer coupled into the fiber (cyan dashed circle), leading to enhanced polarization extinction of $\sim10^7$ in Fig.~\ref{fig:1}d.

To explain these observed intensity distributions, we adapt the theoretical framework developed in Refs.~\cite{bliokh2013goos,bliokh2016spin,benelajla2021physical}. Specifically, we need to first model the spatial field at the focal plane of the objective and then incorporate the single-mode fiber as a confocal Gaussian filter that couples the field into the detector. Details of the calculation are provided in Supplementary Note~\ref{sup_note_2}. Within this framework, the modified Jones matrix of the QWP, which accounts for the finite beam size and spatial filtering, can be expressed in the principal-axis (fast/slow) frame as
\begin{equation}\label{eq:QWPMatrix}
    \bar{\bar{Q}}(x_{plate},y_{plate})=\bar{\bar{Q}}_{0}+\frac{x_{plate}}{2\xi_1}\bar{\bar{Q}}_{1}+\frac{y_{plate}}{2\xi_2}\bar{\bar{Q}}_{2},
\end{equation}
where $x_{plate}$ ($y_{plate}$) denotes the transverse coordinate along the fast (slow) axis of the QWP, and $\xi_1$ and $\xi_2$ are complex fitting parameters when comparing our experimental data. The first term corresponds to the conventional plane-wave Jones matrix of a QWP, 
\changes{
\begin{equation}
    \bar{\bar{Q}}_0
    =e^{-i\Phi_0/2}\begin{bmatrix}1&0\\0&e^{i\Phi_0}\end{bmatrix},
\end{equation}
}where $\Phi_0$ is the phase retardance between the ordinary (o) and extraordinary (e) modes. For an ideal QWP, $\Phi_0=\pi/2$, \changes{which we fix during fitting}. \changes{The second term generates an HG$_{10}$-like contribution along the fast axis,
\begin{equation}
    \bar{\bar{Q}}_{1}=
    \frac{i}{2}\frac{d\Phi_0}{d\theta}e^{-i\Phi_0/2}\begin{bmatrix}-1&0\\0&i\end{bmatrix},
\end{equation}
}\changes{This term is governed by the angular dependence of the retardance. At normal incidence ($\theta=90^\circ$), $\Phi_0$ reaches an extremum and $\frac{d\Phi_0}{d\theta}=0$, so that this first-order contribution vanishes in the ideal case.} The third term generates an HG$_{01}$-like contribution,
\changes{
\begin{equation}
    \bar{\bar{Q}}_{2}
    =-2i\sin(\Phi_0/2)\cot\theta\begin{bmatrix}0&1\\1&0\end{bmatrix},
\end{equation}}\changes{it also vanishes at normal incidence due to $\cot\theta\rightarrow 0$.} Including a rotation $\gamma$ of the QWP fast axis in the transverse plane, the resulting field at the detection plane can be written as
\changes{
\begin{equation}
\label{eq:modelQWP}
\begin{aligned}
\vec E(x_0,y_0)
&\propto 
   \exp\left[-\frac{x_0^{2}+y_0^{2}}{2\omega_f^{2}}\right]\,
   \bar{\bar{A}}(\alpha)\,
   \bar{\bar{Q}}(\gamma)\,
   \bar{\bar{P}}(\beta)\,
   \vec{E}_0.
\end{aligned}
\end{equation}}In Eq.~(\ref{eq:QWPMatrix}), the complex parameters $\xi_1$ and $\xi_2$ characterize the effective amplitude and phase of the first-order spatial contributions along the fast and slow axis directions respectively. For an ideal cylindrically symmetric Gaussian beam, one would expect $\xi_1=\xi_2$. However, imposing this constraint does not fully reproduce the experimental confocal maps. We therefore treat $\xi_1$ and $\xi_2$ as independent complex fitting parameters. \changes{This phenomenological extension indicates an effective transverse anisotropy in the experimental system, while its origin is not uniquely identified in the present work.}

Strictly speaking, in freely propagating paraxial beams, SOI effects are absent unless symmetry is broken \cite{korger2014observation}. \changes{In our system, such symmetry breaking arises mainly from imperfections of the QWP combined with cross-polarized confocal detection. While both $\bar{\bar{Q}}_1$ and $\bar{\bar{Q}}_2$ vanish under ideal normal-incidence conditions, small deviations—such as compound waveplate design, wedge errors, or wavelength mismatch—can introduce a finite effective angular sensitivity, leading to nonzero first-order contributions ($\frac{d\Phi_0}{d\theta}=-0.072$, based on \cite{Thorlabs_QWP}). This explains the experimentally observed two-lobe pattern aligned along the fast axis of the QWP in Fig.~\ref{fig:1}d.}

To get a deeper understanding for the measured modal transformation, we measure and show in Fig.~\ref{fig:2} the evolution of the confocal mappings for different analyzer rotation angles variation $\delta\alpha$ around the symmetrically split mode. \changes{We observe a distinct transition between Gaussian-like modes and first-order HG-like modes, driven by the degree of polarization extinction. Far from maximum extinction, the beam appears as a single Gaussian spot centered on the axis. As the system approaches maximum extinction, this spot shifts away from the center until two distinct lobes emerge, forming an HG$_{10}$-like pattern. Moving past the point of maximum extinction, the original lobe vanishes while the second lobe dominates, eventually shifting back toward the center as the system moves further away from the extinction point.} The fitted confocal maps based on Eq.~(\ref{eq:modelQWP}) show good agreements with our experimental results.  \changes{Although the full model includes both $\xi_1$ and $\xi_2$, the $\bar{\bar{Q}}_2$ contribution is expected to be strongly suppressed in the present normal-incidence geometry. In our case, we find that introducing $\xi_2$ mainly provides additional fitting freedom, with fitted values that fluctuate substantially across datasets and do not exhibit clear physical robustness. We therefore exclude $\xi_2$ from the final fitting and retain only $\xi_1$, leading to a more compact model without significant loss of fitting quality. The fitted $\xi_1$ is around $(0.705-0.406\mathrm{i})$ $\mu$m and the corresponding root-mean-square deviation (RMSD) are listed in Supplementary Table~\ref{tab:AnalyzerDependent}. Notably, $\xi_1$ appears in the denominator of the first-order term, so a smaller $|\xi_1|$ corresponds to a stronger HG$_{10}$-like contribution. In our fits, the magnitude of $\xi_1$ is much smaller than the focal length $f=26$~mm, which would normally be expected to set the theoretical scale of $\xi_1$ as derived in Supplementary Note~\ref{sup_note_2}. This indicates that the observed lobe intensities are substantially stronger than predicted. What's more, for each analyzer setting, we extract the positions of the dominant peaks, showing a clear anti-crossing behavior in Supplementary Fig.~\ref{fig:S3}, similar to that reported in Ref.~\cite{benelajla2021physical}.}

\textbf{Polarization-induced reorientation of first-order HG-like modes.}
Finally, we discuss the tunability of the transverse lobe orientation under cross-polarized detection in Fig.~\ref{fig:3}. We investigate how the fast-axis orientation of the QWP in the $x$–$y$ plane, characterized by the angle $\gamma$, influences the observed two-lobe intensity patterns. To achieve an incident linear polarization, the input polarization is aligned either parallel (solid frames) or perpendicular (dotted frames) to the fast axis of the QWP.

As shown in Fig.~\ref{fig:3}, \changes{rotating the fast axis of the QWP leads to a corresponding rotation of the observed two-lobe pattern. Notably, the orientation of this pattern is not confined to the transverse coordinate axes, as would be expected for pure HG$_{01}$ or HG$_{10}$ modes. Instead, by adjusting $\gamma$, the pattern can be reoriented, including along the $+45^\circ$ and $-45^\circ$ directions.} This behaviour is naturally understood as a polarization-dependent reorientation of the transverse field distribution. Within our model, the rotation $\bar{\bar{Q}}(\gamma)$ in Eq.~(\ref{eq:modelQWP}) redistributes the first-order HG$_{10}$-like spatial contributions, thereby providing continuous control over the lobe orientation. A similar reorientation is observed when varying the angle of incidence on the QWP, as shown in Supplementary Fig.~\ref{fig:S4}.

\changes{To further confirm that the observed SOI is not a detection-filtering artifact, we show in Supplementary Fig.~\ref{fig:S5} direct imaging of the beam profile under cross-polarization, recorded by a camera placed immediately after the analyzer in the collimated beam. The images in Fig.~\ref{fig:S5}b clearly reveal the two-lobe pattern for different polarizer and QWP configurations. This confirms that the observed two-lobe structure is a clear signature of spin-orbit interaction induced by the QWP.}

\changes{As a final note, we observe a small but systematic offset of the geometric center of the two-lobe pattern with respect to the center defined by the co-polarized reference, indicated by the dashed white lines in Fig.~\ref{fig:3}. 
The overall displacement of the two-lobe pattern may be related to polarization-mixing-induced transverse shifts in the QWP~\cite{mazanov2020photonic}. In addition, the orientation of the two-lobe pattern varies systematically with the QWP fast-axis angle, as summarized in Fig. \ref{fig:S6}. The extracted lobe-axis angle exhibits an approximately linear dependence on the QWP fast-axis orientation, in overall agreement with the model, confirming the continuous tunability of the lobe rotation.}


\changes{From an application perspective, our results clarify the role of the QWP in effectively enhancing polarization extinction in confocal systems, which is particularly relevant for resonance measurements in epifluorescence geometries where suppressing parasitic laser background is critical. Furthermore, the demonstrated polarization-dependent reorientation of first-order HG-like modes provides a simple and robust approach to transverse beam shaping using only standard polarization optics. Unlike approaches based on spatial light modulators \cite{forbes2016creation} or interferometric mode converters \cite{ishaaya2003conversion}, the present scheme enables continuous control of structured focal fields within a compact and alignment-tolerant confocal geometry. This capability may find applications in polarization-resolved confocal microscopy \cite{kuhlmann2013dark}, mode-selective coupling into anisotropic waveguides \cite{martin2025purcell} or optical fibers \cite{li2023metafiber}, and spatially structured excitation of quantum emitters or nanostructures \cite{montagnac2023control}. More broadly, our results highlight that spatial degrees of freedom can be engineered through polarization control even in nominally paraxial and collimated systems, suggesting new design strategies for polarization-based mode control in both free-space and integrated platforms.}

\changes{While our current analysis provides a qualitative framework for the observed phenomena, a significant quantitative paradox remains. Both the central mirror-based measurements \cite{benelajla2021physical} and the QWP experiments reveal a clear modal transformation from a Gaussian to a HG$_{10}$ or HG$_{01}$ mode under maximum extinction. However, when interpreted through the lens of standard Fourier optics, both experimental sets fail to account for the lobe intensities, which are measured to be several orders of magnitude stronger than theory predicts. This systematic discrepancy across two distinct experimental setups suggests that the symmetry breaking leading to modal transformation involves a more fundamental mechanism than is currently modeled. Whether we are overlooking a subtle classical effect or if the confocal arrangement is revealing a systematically new aspect of light-matter interaction remains an open question. Rather than forcing a quantitative fit, we present this discrepancy as a central challenge; resolving why these mode intensities are so unexpectedly robust will be essential for a complete rethinking of the physics within these optical systems.}

\textbf{Acknowledgments.} \changes{We thank Daria Markina, Sai Shradha and Markus Lohnes for technical assistance. W.L. acknowledges Thorlabs Inc. for providing the vendor data used in this work.}

\textbf{Author Contributions.} 
B.U. conceived and supervised the project. W.L. designed the experiments and performed the optical measurements. W.L. and A.L. carried out the numerical simulations. A.L. interfaced the experimental setup. W.L. and A.L. developed the theoretical framework, building upon previous work by K.K., M.B. Data analysis was performed by W.L., A.L., and B.U. All authors discussed the results. W.L., A.L., and B.U. wrote the manuscript with input from all authors.

\textbf{Data Availability.} 
The data that support the findings of this study are available from the corresponding
authors upon request.

\textbf{Competing interests.} The authors declare no competing interests.

\clearpage
\onecolumngrid

\section*{Supplementary Materials for\\ 
``\textit{Role of a Quarter-Wave Plate in Confocal Microscopy: Signature of Spin-Orbit Interactions}"}
\setcounter{figure}{-1}
\renewcommand{\thefigure}{S\arabic{figure}}

\setcounter{table}{-1}
\renewcommand{\thetable}{S\arabic{table}}

\setcounter{section}{0}
\renewcommand{\thesection}{S\arabic{section}}

\tableofcontents

\section*{Methods}
\subsection*{Experimental setup}
\noindent
We implement a compact confocal optical configuration, illustrated in Fig.~\ref{fig:1}a of the main text, to isolate the physical mechanism responsible for the enhanced laser rejection. 
A diode laser operating at $\lambda=850$~nm is launched into a single-mode fiber, whose angled ($4^\circ$) flat-polished output produces a nearly diffraction-limited Gaussian beam with a mode-field radius $\omega_0=2.5~\mu$m (defined at the $1/e^2$ intensity level).
The diverging beam from the fiber is collimated using a diffraction-limited microscope objective (NA = 0.25, $f=26$~mm), resulting in a Gaussian beam with a waist radius of approximately 3~mm. 
The numerical aperture of the objective is chosen to exceed the divergence half-angle of the fiber mode in order to preserve the spatial Gaussian profile during collimation.
The collimated beam is then transmitted through a high-extinction nanoparticle thin-film linear polarizer (LPVIS050, Thorlabs GmbH) mounted on a piezoelectric rotation stage (ECR4040/Al/NUM/RT, attocube systems AG) providing an angular resolution of 0.001$^\circ$. 
After polarization preparation, the beam passes through a zero-order quartz quarter-wave plate (WPQ05M-850, Thorlabs GmbH), which serves as the central element of the experiment. 
A zero-order design is selected to minimize wavelength and temperature sensitivity compared with multi-order retarders. 
The wave plate is mounted in a precision rotation mount (CRM1PT/M, Thorlabs GmbH) with a Vernier resolution of approximately $0.083^\circ$.
For reference measurements, the quarter-wave plate can be removed from the optical path. 
The beam subsequently passes through a second linear polarizer acting as an analyzer, identical to the input polarizer and mounted on the same type of piezoelectric rotation stage.
Finally, the transmitted light is refocused into a second single-mode fiber using a microscope objective identical to the collimating lens, providing spatial-mode filtering in a confocal detection geometry. The collecting fiber is mounted on a three-axis translation stage (MDE330TH, Thorlabs GmbH) equipped with both piezo actuators and manual adjustment screws (PE4, Thorlabs GmbH). The optical signal is detected at the fiber output using a photodiode (OE-200-SI-FC, FEMTO Messtechnik GmbH). This signal is read out by a multimeter (Keysight 34465A 6,5 Digit Multimeter) connected to a computer, interfacing with all devices and coordinating the measurement process. The optimization is also automated using a Python control routine that iteratively adjusts the polarizer and analyzer angles to maximize the extinction ratio, with a minimum angular step of $\sim0.005^\circ$ for the polarizer.

\section{Supplementary Notes 1: Plane-wave Jones-matrix analysis for different geometries}
\label{sup_note_1}
\noindent
For clarity, we first define the Jones matrices used throughout the following derivations.
\noindent
A linear polarizer aligned along the $x$ axis with finite leakage $b$ is written as
\begin{equation}
\bar{L}\bar{P}_{p_0}=
\begin{bmatrix}
a & 0\\
ib & 0
\end{bmatrix},
\end{equation}
where $a^2+b^2=1$, $a,b\in\mathbb{R}$, and $a^2\gg b^2$.
\noindent
The rotation matrix is
\begin{equation}
\bar{R}(\varphi)=
\begin{bmatrix}
\cos\varphi & -\sin\varphi\\
\sin\varphi & \cos\varphi
\end{bmatrix}.
\end{equation}
\noindent
The polarizer rotated by an angle $\beta$ is
\begin{equation}
\bar{P}(\beta)=\bar{R}(\beta)\bar{L}\bar{P}_{p_0}\bar{R}(-\beta),
\end{equation}
and the analyzer rotated by $\alpha$ is
\begin{equation}
\bar{A}(\alpha)=\bar{R}(\alpha)\bar{L}\bar{A}_{p_0}\bar{R}(-\alpha).
\end{equation}
\noindent
The zero-order half-wave plate (fast axis along $x$) is
\begin{equation}
\bar{H}=-i
\begin{bmatrix}
1 & 0\\
0 & -1
\end{bmatrix},
\end{equation}
and the zero-order quarter-wave plate is
\begin{equation}
\bar{Q}=e^{-i\pi/4}
\begin{bmatrix}
1 & 0\\
0 & i
\end{bmatrix}.
\end{equation}
\noindent
Unless otherwise specified, the incident field is
\begin{equation}
\vec{E}_0=
\begin{bmatrix}
1\\
0
\end{bmatrix}.
\end{equation}

\subsection{Geometry 1: \textit{Polarizer–Quarter-Wave Plate–Analyzer}}
\noindent
The input field is rotated by $\beta$ to align with the polarizer to get the maximum signal,
\begin{equation}
\vec{E}(\beta)=\bar{R}(\beta)\vec{E}_0.
\end{equation}
\noindent
After passing through the polarizer, quarter-wave plate, and analyzer, the field becomes
\begin{equation}
\vec{E}=\bar{A}(\alpha)\bar{Q}\bar{P}(\beta)\bar{R}(\beta)\vec{E}_0.
\end{equation}
\noindent
Straightforward calculation gives the detected intensity
\begin{equation}
I \propto\vec{E}^{\,*}\cdot\vec{E}= \cos ^2\beta (a \cos \alpha -b \sin \alpha )^2+\sin ^2\beta  (b \cos \alpha -a \sin \alpha )^2
\end{equation}

\noindent
These equations admit solutions for $\alpha$ and $\beta$, demonstrating that the insertion of a quarter-wave plate enables complete cancellation of the leakage field.
\noindent
For example, fixing $\beta=0$ yields
\begin{equation}
\boxed{
\cos^2\alpha=b^2.}
\end{equation}


\subsection{Geometry 2: \textit{Polarizer–Analyzer}}
\noindent
Without additional polarization elements,
\begin{equation}
\vec{E}=\bar{A}(\alpha)\bar{P}(\beta)\bar{R}(\beta)\vec{E}_0.
\end{equation}
\noindent
The detected intensity becomes
\begin{equation}
I \propto\vec{E}^{\,*}\cdot\vec{E}=
a^2\cos^2(\alpha-\beta)+b^2\sin^2(\alpha-\beta).
\end{equation}
\noindent
Perfect extinction would require
\begin{equation}
\cos^2(\alpha-\beta)=\frac{-b^2}{a^2-b^2},
\end{equation}
which has no physical solution because the left-hand side is non-negative. 
Therefore, a polarizer-analyzer pair alone cannot eliminate the leakage field.
\noindent
For $\beta=0$ and $\alpha=\pi/2$,
\begin{equation}
I=b^2.
\end{equation}

\subsection{Geometry 3: \textit{Polarizer–Half-Wave Plate–Analyzer}}
\noindent
With a half-wave plate inserted between polarizer and analyzer,
\begin{equation}
\vec{E}=\bar{A}(\alpha)\bar{H}\bar{P}(\beta)\bar{R}(\beta)\vec{E}_0.
\end{equation}
\noindent
The detected intensity becomes
\begin{equation}
I \propto \vec{E}^{\,*}\cdot\vec{E}=a^2\cos^2(\alpha+\beta)+b^2\sin^2(\alpha+\beta).
\end{equation}
\noindent
No choice of $\alpha$ and $\beta$ can make the intensity vanish, showing that a half-wave plate alone does not enable complete extinction of the leakage field.
\noindent
For $\beta=0$ and $\alpha=\pi/2$,
\begin{equation}
I=b^2.
\end{equation}

\newpage
\section{Supplementary Notes 2: Derivation of the Intensity collected by the Single Mode Fiber}
\label{sup_note_2}
\subsection{Modified Jones matrix for quarter-wave plate}
\begin{figure*}[htbp]
	\centering
	\includegraphics[width=0.5\textwidth]{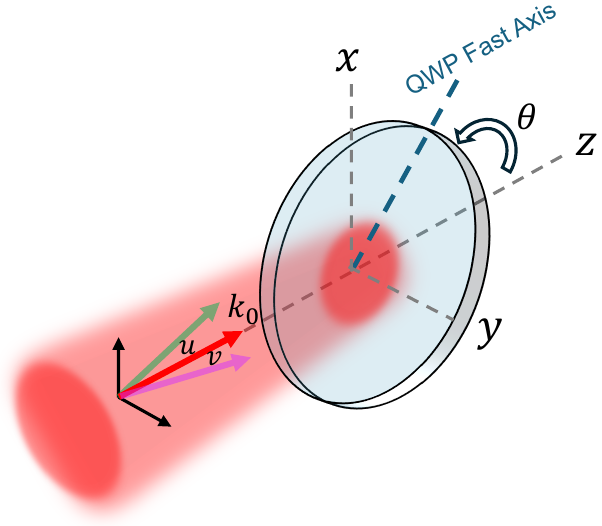}
	\caption{Schematic of a finite-size Gaussian beam incident on a quarter-wave plate (QWP). The central wavevector is denoted by $k_0$, while $u$ and $v$ represent the angular coordinate of the constituent plane-wave components. The fast axis of the QWP is oriented at an angle $\theta$ with respect to the $z$ axis.}
	\label{fig:S0}
\end{figure*}

\noindent
Let us consider the general case for transmission of a Gaussian beam through a thin positive uniaxial birefringent crystal plate (e.g., quartz). The beam propagates along the $z$ axis, while the fast axis of the plate lies in the $x$–$z$ plane at an angle $\theta$ with respect to the $z$ axis (Fig.~\ref{fig:S0}).

\noindent
We consider a polarized paraxial Gaussian beam propagating along the $z$-axis in free space. The beam represents a superposition of multiple plane waves with close wave vectors $\vec{k}=k_p \vec{p}+k_s \vec{s}+k_z \vec{k}_z$ with wave number $k_0=2\pi/\lambda$. If we define the angular coordinates $u=\frac{k_p}{k_0}$, $v=\frac{k_s}{k_0}$, beam radius $\omega_0$, and beam divergence $\theta_0=\frac{2}{k_0\omega_0}$, then the normalized angular distribution at the beam waist ($z=0$) is
\begin{equation}\label{5}
    \hat{\mathcal{E}}(k_p,k_s,0)=\frac{\omega_0^2}{4\pi}e^{-(k_p^2+k_s^2)\frac{\omega_0^2}4}\vec{E}_0=\frac{\omega_0^2}{4\pi}e^{-\frac{k_p^2+k_s^2}{k_0^2\theta_0^2}}\vec{E}_0=\frac{\omega_0^2}{4\pi}e^{-\frac{u^2+v^2}{\theta_0^2}}\vec{E}_0.
\end{equation}
Under paraxial approximation ($k_p,k_s\ll|k_0|$):
\begin{equation}\label{2}
    k_z=k_0\sqrt{1-\frac{k_p^2+k_s^2}{k_0^2}}\cong k_0\left(1-\frac{k_p^2+k_s^2}{2k_0^2}\right).
\end{equation}
\noindent
For simplicity, suppose the waist plane of incident Gaussian beam is at the surface of the plate. It is well known that the crystal plate induces linear birefringence between the ordinary ($o$) and extraordinary ($e$) polarization modes, which propagate with slightly different phase velocities. For the central plane wave in the beam and a not rotated QWP, the $e$ and $o$ modes correspond to the $x-$ and $y-$ linear polarizations. In practice, one would also need to introduce an additional basis transformation to account for the rotation of the QWP around the z-axis. Thus, the action of the plate on the central plane wave can be characterized by the standard Jones equation and matrix:
\begin{equation}
    \hat{\mathcal{E}}^\prime(0,0,0)=\bar{\bar{Q}}_0\hat{\mathcal{E}}(0,0,0),
\end{equation}
\begin{equation}\label{0oder}
    \bar{\bar{Q}}_0=\begin{bmatrix}e^{-i\Phi_0/2}&0\\0&e^{i\Phi_0/2}\end{bmatrix},
\end{equation}
where $\Phi_0$ is the phase difference between the $o$ and $e$ modes, which is acquired upon the propagation in the plate, $\hat{\mathcal{E}}^\prime(0,0,0)$ indicates the field of the transmitted central wave, and we ignore the common phase factor. Specifically, for quarter-wave plate, $\Phi_0=\pi/2$.

\noindent
Importantly, the transmission Jones matrix in Eq.~(\ref{0oder}) strictly describes only the central plane-wave component of the beam (corresponding to zero angular deviation). 
A realistic beam, however, contains an angular spectrum of Fourier components that propagate along slightly different directions and therefore experience slightly different polarization transformations inside the anisotropic element. 
Detailed derivation can be found in refs \cite{bliokh2013goos,bliokh2016spin}. Taking these first-order angular corrections into account leads to a transmission Jones matrix that depends explicitly on the angular coordinates $(u,v)$:

\begin{equation}
    \hat{\mathcal{E}}^\prime(u,v,0)=\bar{\bar{Q}}(u,v)\hat{\mathcal{E}}(u,v,0),
\end{equation}
\begin{equation}
    \bar{\bar{Q}}(u,v)\simeq \begin{bmatrix}
        e^{-i\Phi_0/2}(1+u\mathcal{X}_e)&e^{-i\Phi_0/2}v\mathcal{Y}_e\\
        -e^{i\Phi_0/2}v\mathcal{Y}_o&e^{i\Phi_0/2}(1+u\mathcal{X}_o)
    \end{bmatrix},
\end{equation}
where
\begin{equation}
    \mathcal{X}_{e,o}=\mp\frac{i}{2}\frac{d\Phi_0}{d\theta},
\end{equation}
\begin{equation}
    \mathcal{Y}_{e,o}=[1-\exp(\pm i\Phi_0)]\cot\theta.
\end{equation}
To express the result conveniently, we split the modified matrix $\bar{\bar{Q}}(u,v)$ into three terms: $\bar{\bar{Q}}(u,v)=\bar{\bar{Q}}_0+u\bar{\bar{Q}}_{1}+v\bar{\bar{Q}}_{2}$, where
\begin{equation}
    \bar{\bar{Q}}_0=\begin{bmatrix}e^{-i\Phi_0/2}&0\\0&e^{i\Phi_0/2}\end{bmatrix},
\end{equation}
\begin{equation}
    \bar{\bar{Q}}_{1}=\begin{bmatrix}e^{-i\Phi_0/2}\mathcal{X}_e&0\\0&e^{i\Phi_0/2}\mathcal{X}_o\end{bmatrix},
\end{equation}
\begin{equation}
    \bar{\bar{Q}}_{2}=\begin{bmatrix}0&e^{-i\Phi_0/2}\mathcal{Y}_e\\-e^{i\Phi_0/2}\mathcal{Y}_o&0\end{bmatrix}.
\end{equation}

\subsection{Spatial field distribution at a distance $z$ after transmission from the quarter-wave plate}
\noindent
In angular spectrum representation, the spatial distribution $\vec{E}(p,s,z)$ is the two-dimensional Fourier transform of the angular distribution $\hat{\mathcal{E}}(k_p,k_s,0)$:
\begin{equation}\label{1}
    \vec{E}(p,s,z)=\iint_{-\infty}^{+\infty}e^{+\mathrm{i} k_pp+\mathrm{i} k_ss+\mathrm{i} k_zz}\hat{\mathcal{E}}(k_p,k_s,0) \mathrm{d}k_p\mathrm{d}k_s.
\end{equation}
Substitute Eq.(\ref{2}) to Eq.(\ref{1}), then
\begin{equation}\label{3}
    \vec{E}(p,s,z)\cong\iint_{-\infty}^{+\infty}e^{+ik_pp+ik_ss+ik_0\left(1-\frac{k_p^2+k_s^2}{2k_0^2}\right)z}\hat{\mathcal{E}}(k_p,k_s,0)dk_pdk_s.
\end{equation}
After transmitted through the wave plate, the field becomes
\begin{equation}\label{6}
    \hat{\mathcal{E}}^\prime(k_p,k_s,0)=\frac{\omega_0^2}{4\pi}e^{-\frac{\omega_0^2(k_p^2+k_s^2)}4}\left\{\bar{\bar{Q}}_0+u\bar{\bar{Q}}_{1}+v\bar{\bar{Q}}_{2}\right\}\vec{E}_0,
\end{equation}
i.e., 
\begin{equation}\label{7}
    \hat{\mathcal{E}}^\prime(k_p,k_s,0)=\frac{\omega_0^2}{4\pi}e^{-\frac{\omega_0^2(k_p^2+k_s^2)}4}\left\{\bar{\bar{Q}}_0+\frac{k_p}{k_0}\bar{\bar{Q}}_{1}+\frac{k_s}{k_0}\bar{\bar{Q}}_2\right\}\vec{E}_0.
\end{equation}
The Fourier transform we need to perform is
\begin{equation}\label{8}
    \hat{\mathcal{E}}^{\prime}(k_p,k_s,k_z)=e^{+i\left(k_0-\frac{k_p^2+k_s^2}{2k_0}\right)z}\hat{\mathcal{E}}^\prime(k_p,k_s,0).
\end{equation}
Substitute Eq.(\ref{7}) to Eq.(\ref{8}), then
\begin{equation}\label{9}
    \hat{\mathcal{E}}^{\prime}(k_p,k_s,k_z)=\frac{\omega_0^2}{4\pi}e^{+ik_0z-\frac{\omega_0^2}4(1+i\frac z{k_0\omega_0^2/2})(k_p^2+k_s^2)}\left\{\bar{\bar{Q}}_0+\frac{k_p}{k_0}\bar{\bar{Q}}_{1}+\frac{k_s}{k_0}\bar{\bar{Q}}_2\right\}\vec{E}_0.
\end{equation}
Rayleigh length is defined by
\begin{equation}\label{10}
    \ell=k_0\omega_0^2/2.
\end{equation}
Also, define the notation
\begin{equation}\label{12}
    \omega_z^2=\omega_0^2(1+iz/\ell), \ell_z=\ell+iz,
\end{equation}
then the Eq.(\ref{9}) becomes
\begin{equation}\label{11}
    \hat{\mathcal{E}}^{\prime}(k_p,k_s,k_z)=\frac{\omega_0^2}{4\pi}e^{+ik_0z}e^{-\frac{\omega_z^2(k_p^2+k_s^2)}4}\left\{\bar{\bar{Q}}_0+\frac{k_p}{k_0}\bar{\bar{Q}}_{1}+\frac{k_s}{k_0}\bar{\bar{Q}}_{2}\right\}\vec{E}_0.
\end{equation}
\noindent
The spatial distribution of the field at a distance $z$ of the wave plate (and before entering the objective) is
\begin{equation}\label{13}
\begin{split}
\vec{E}(p,s,z)
&=\iint_{-\infty}^{+\infty}\hat{\mathcal{E}}^{\prime}\left(k_p,k_s,k_z\right){e^{ik_pp}e^{ik_ss}}dk_pdk_s\\
&=\frac{\omega_0^2}{4\pi}e^{+ik_0z}\iint_{-\infty}^{+\infty}e^{-\frac{\omega_z^2(k_p^2+k_s^2)}4}\left\{\bar{\bar{Q}}_0+\frac{k_p}{k_0}\bar{\bar{Q}}_{1}+\frac{k_s}{k_0}\bar{\bar{Q}}_{2}\right\}\vec{E}_0e^{ik_pp}e^{ik_ss}dk_pdk_s\\
&=\frac{\omega_0^2}{4\pi} \frac{4\pi}{\omega_z^2} e^{+ik_0z-\frac{p^2+s^2}{\omega_z^2}}\left\{\bar{\bar{Q}}_0+\frac{ip}{k_0\omega_z^2/2}\bar{\bar{Q}}_{1}+\frac{is}{k_0\omega_z^2/2}\bar{\bar{Q}}_{2}\right\}\vec{E}_0\\
&=\boxed{\frac{\omega_0^2}{\omega_z^2}e^{ik_0z-\frac{p^2+s^2}{\omega_0^2(1+iz/\ell)}}\left\{\bar{\bar{Q}}_0+\frac{ip}{\ell+iz}\bar{\bar{Q}}_{1}+\frac{is}{\ell+iz}\bar{\bar{Q}}_{2}\right\}\vec{E}_0}\\
&=\boxed{\frac{\omega_0^2}{\omega_z^2}e^{ik_0z-\frac{p^2+s^2}{\omega_z^2}}\Bigl\{\bar{\bar{Q}}_0+\frac{ip}{\ell_z}\bar{\bar{Q}}_{1}+\frac{is}{\ell_z}\bar{\bar{Q}}_{2}\Bigr\}\vec{E}_0}\\
\end{split}
\end{equation}

\subsection{Spatial field distribution at focal point of the lens}
\noindent
In the aforementioned section, we have calculated the spatial distribution transmitted after the plate, the next step is to utilize the Fourier transforming properties of lens to get the spatial field distribution at focal point of the lens.
\noindent
If $z$ is the distance from the wave plate to the lens itself:
\begin{equation}\label{14}
    \vec{E}(x,y,z)=\frac{-i}{\lambda f}e^{\frac{ik_0(x^2+y^2)}{2f}}\iint_{-\infty}^{+\infty}\vec{E}(p,s,z)e^{-i\left(\frac{k_0x}f\right)p}e^{-i\left(\frac{k_0y}f\right)s}dsdp.
\end{equation}
\noindent
If $z$ is the distance from the wave plate to the back focal plane of the lens:
\begin{equation}\label{15}
    \vec{E}(x,y,z)=-\frac i{\lambda f}\iint_{-\infty}^{+\infty}\vec{E}(p,s,z)e^{-i\left(\frac{k_0x}f\right)p}e^{-i\left(\frac{k_0y}f\right)s}dsdp.
\end{equation}
Making use of $\ell_{z}=\ell(1+iz/\ell)$, $\begin{aligned}\omega_z^2=\omega_0^2(1+iz/\ell)\end{aligned}$, and $k_0\omega_0^2/2=\ell $, the spatial field distribution at the focal point is
\begin{equation}\label{17}
\begin{split}
    \vec{E}(x,y,f)
    &=-\frac{ie^{ik_0z}}{\lambda f}\frac{\omega_0^2}{\omega_z^2}\iint_{-\infty}^{+\infty}\left\{\bar{\bar{Q}}_0+\frac{ip}{\ell_z}\bar{\bar{Q}}_{1}+\frac{is}{\ell_z}\bar{\bar{Q}}_{2}\right\}\vec{E}_0e^{-\frac{p^2+s^2}{\omega_z^2}}e^{-i(\frac{k_0x}f)p}e^{-i(\frac{k_0y}f)s}dsdp\\
    &=-\frac{ie^{ik_0z}}{\lambda f}\frac{\omega_0^2}{\omega_z^2}e^{-\frac{\omega_{z}^{2}k_{0}^{2}(x^{2}+y^{2})}{4f^{2}}}\pi\omega_z^2\left\{\bar{\bar{Q}}_0+\frac{k_0\omega_z^2x}{2f\ell_z}\bar{\bar{Q}}_{1}+\frac{k_0\omega_z^2y}{2f\ell_z}\bar{\bar{Q}}_{2}\right\}\vec{E}_0\\
\end{split}
\end{equation}
\noindent
We can verify that
\begin{equation}\label{18}
    \frac{\omega_z^2k_0^2(x^2+y^2)}{4f^2}=(1+iz/\ell)\frac{x^2+y^2}{\omega_f^2}, \frac{k_0\omega_z^2}{2\ell_z}=1
\end{equation}
Substitute Eq.(\ref{18}) to Eq.(\ref{17}) and use $\lambda f=\pi\omega_0\omega_f$, then
\begin{equation}\label{19}
    \boxed{\vec{E}(x,y,f)=-i\frac{\omega_0}{\omega_f}e^{ik_0z-(1+iz/\ell)\frac{x^2+y^2}{\omega_f^2}}\left\{\bar{\bar{Q}}_0+\frac xf\bar{\bar{Q}}_{1}+\frac yf\bar{\bar{Q}}_{2}\right\}\vec{E}_0}
\end{equation}

\subsection{Single-mode fiber collection and convolution}
\noindent
The single mode optical fiber has a normalized filter function
\begin{equation}\label{20}
   \vec{E}_{filter}=\frac{1}{\pi\omega_f^2}e^{-\frac{x_0^2+y_0^2}{\omega_f^2}}
\end{equation}
The convolution of the field with this filter function provides the field that is collected into the fiber and ported to the detector
\begin{equation}\label{21}
    \vec{E}(x_0,y_0,f)=\vec{E}(x,y,f)*\vec{E}_{filter}=\frac1{\pi\omega_f^2}\iint_{-\infty}^{+\infty}e^{-\frac{(x-x_0)^2+(y-y_0)^2}{\omega_f^2}}\vec{E}(x,y,f)dxdy
\end{equation}
\noindent
Carrying out the Gaussian integrals yields the following closed form for the coupled field:
\begin{equation}
    \vec{E}(x_0,y_0,f)=\frac{-i}{2 \xi_0} \frac{\omega_0}{\omega_f}e^{ik_0z-\gamma\frac{x_0^2+y_0^2}{2\omega_f^2}}\bar{\bar{A}}(\alpha)\left\{\bar{\bar{Q}}_0+\frac{x_0}{2\xi_1}\bar{\bar{Q}}_{1}+\frac{y_0}{2\xi_2}\bar{\bar{Q}}_{2}\right\}\bar{\bar{P}}(\beta)\vec{E}_0,
\end{equation}
with the auxiliary quantities
\[
    \xi_0=\left(1+\frac{iz}{2\ell}\right),\qquad
    \xi_{1,2}= f \left(1+\frac{iz}{2\ell}\right),\qquad
    \gamma=\frac{1+iz/\ell}{1+iz/(2\ell)}.
\]
Since in our setup the propagation distance \(z\) between the wave plate and the back focal plane is much smaller than the Rayleigh range \(\ell\), we can apply the approximation \(z\ll\ell\) to simplify these auxiliary quantities. Under this limit,
\[
    \xi_0 \to 1, \qquad
    \xi_{1,2} \to f ,
\]
which would remove the associated phase and amplitude corrections. However, performing this simplification would make the theoretical model unable to reproduce the experimental data.

\clearpage
\section{Supplementary Data}
\subsection{Effect of $\Phi_0$ on the extinction ratio}
\begin{figure}[htbp]
	\centering
	\includegraphics[width=1.0\textwidth]{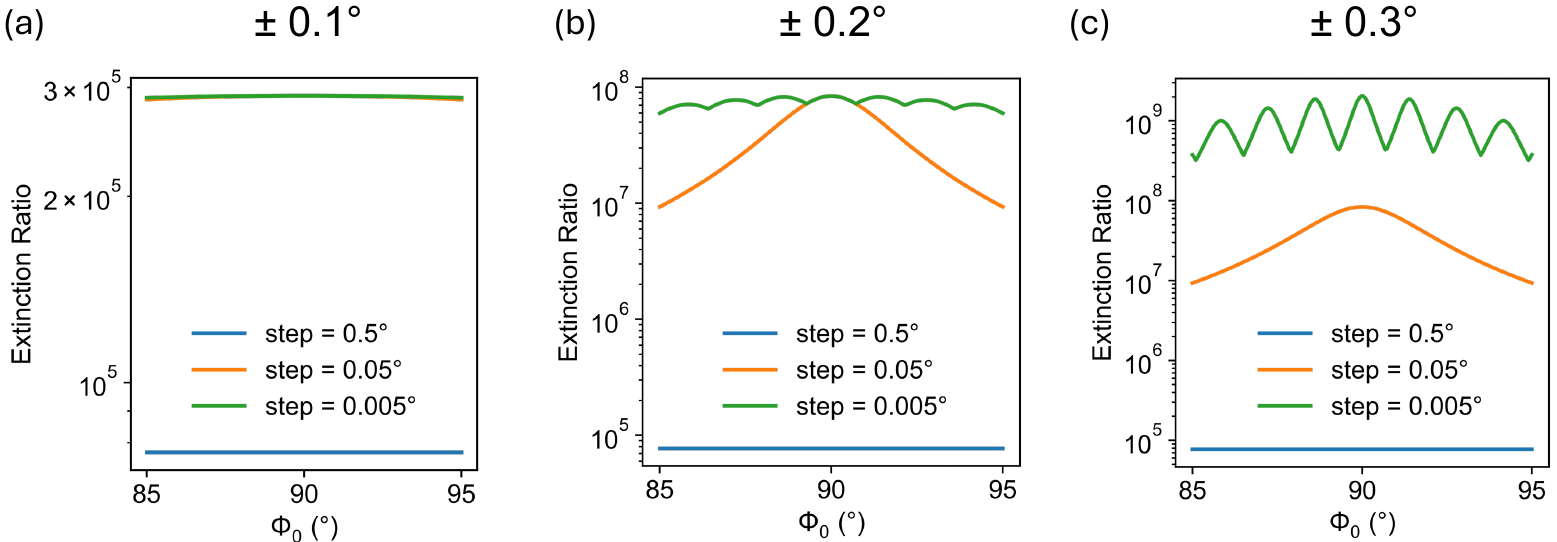}
	\caption{Optimized extinction ratio as a function of the QWP retardance $\Phi_0$ for different angular tuning windows of the polarizer and analyzer: (a) $\pm0.1^\circ$, (b) $\pm0.2^\circ$, and (c) $\pm0.3^\circ$. Within each window, the optimization is performed on a discrete angular grid, where the angular step size $\Delta\alpha=\Delta\beta$ is indicated in the legend. 
    The finite tuning window and discrete angular sampling impose a practical upper bound on the achievable extinction ratio, which is on the order of $10^7$ for the experimentally relevant window of $\pm0.2^\circ$.}
    \label{fig:S1}
\end{figure}

\noindent
In practice, the maximum extinction is obtained by discretely optimizing the polarizer and analyzer angles within a finite tuning window and with a finite angular step size, both determined by the resolution of the rotators. To quantify this effect, we numerically evaluate the optimized extinction ratio as a function of the QWP retardance $\Phi_0$ for different angular tuning windows.
\noindent
In the optimization, the polarizer and analyzer angles are restricted to a window of $\pm 0.2^\circ$ around the cross-polarization condition, and the angular step size is set to $\Delta\beta\approx 0.005^\circ$, consistent with the experimental optimization procedure. The resulting extinction ratio is shown in Fig.~\ref{fig:S1} (b). Due to the finite tuning window and discrete angular sampling, the achievable extinction ratio is typically limited to $[10^7,10^8]$, even when $\Phi_0$ is close to the ideal quarter-wave retardance.
\noindent
This practical limit becomes more restrictive for smaller angular tuning windows as shown in Fig.~\ref{fig:S1} (a), while larger windows in Fig.~\ref{fig:S1} (c) allow higher extinction ratios but introduce oscillatory behavior associated with discrete angular sampling. In addition, any wavelength mismatch between the laser and the QWP retardance further reduces the achievable extinction ratio.

\newpage
\subsection{Modal transformation and extinction ratio curve for ``polarizer-half wave plate (HWP)-analyzer" geometry}
\noindent
In the \textit{polarizer-half wave plate (HWP)-analyzer} geometry, we did the same measurement under co- and cross-polarization. The extinction ratio is limited to $\sim10^6$. Both the confocal mapping under co- and cross-polarization show Gaussian-like mode pattern.
\begin{figure}[htbp]
	\centering
	\includegraphics[width=0.85\textwidth]{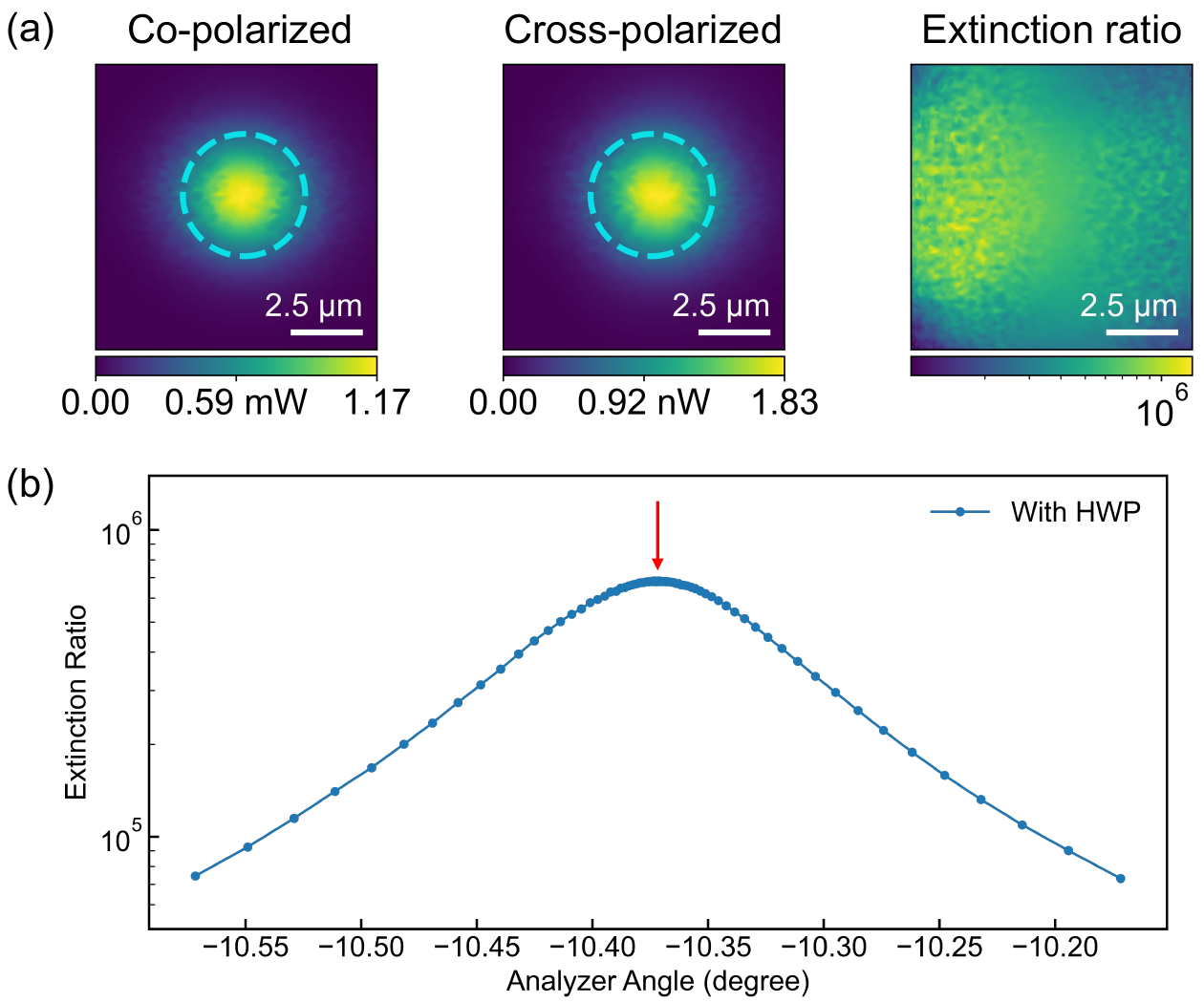}
	\caption{Modal transformation (a) and extinction ratio curve (b) for polarizer-half wave plate-analyzer geometry. Red arrow in (b) indicates the analyzer angle that we used to measure the cross-polarized modal mapping in (a).}
	\label{fig:S2}
\end{figure}

\newpage
\subsection{Anti-crossing behavior of the mode peak}
\noindent
\changes{Here we extract the peak positions as a function of analyzer offset $\delta\alpha$ from the data shown in Fig.~\ref{fig:2} of the main text. For each analyzer setting, the positions of the dominant peaks are identified and plotted versus $\delta\alpha$. The two branches approach each other near $\delta\alpha=0^\circ$ but remain separated, showing a clear anti-crossing behavior, similar to that reported in Ref.~\cite{benelajla2021physical}.}
\begin{figure}[htbp]
	\centering
	\includegraphics[width=0.55\textwidth]{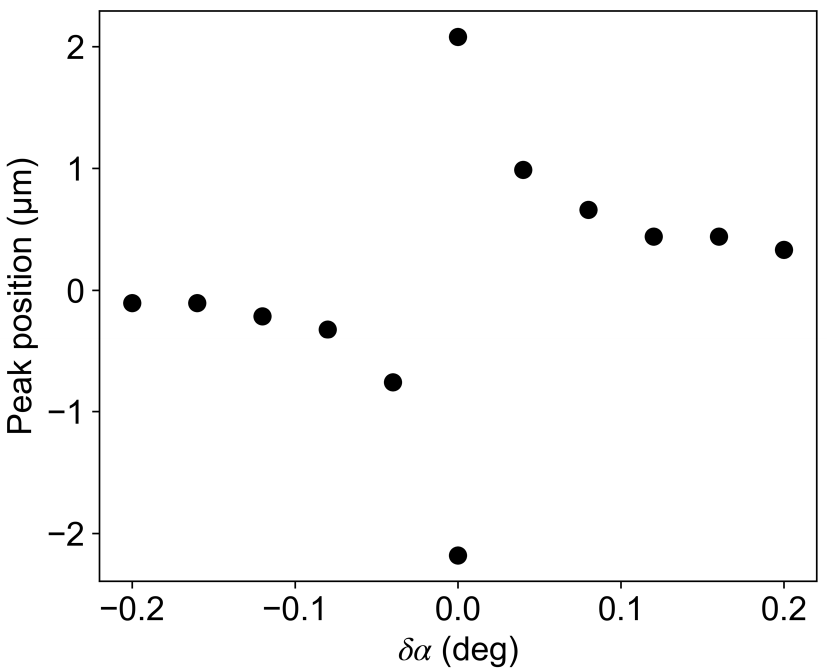}
	\caption{\changes{Analyzer-dependent evolution of the extracted peak positions from the data in Fig.~\ref{fig:2} of the main text. The peak positions of the two dominant branches are plotted as functions of analyzer offset $\delta\alpha$. An avoided crossing is observed near $\delta\alpha=0^\circ$, where the two branches come close but do not intersect.
}}
	\label{fig:S3}
\end{figure}

\newpage
\subsection{Modal transformation under different angle-of-incidence}
\begin{figure}[htbp]
	\centering
	\includegraphics[width=0.85\textwidth]{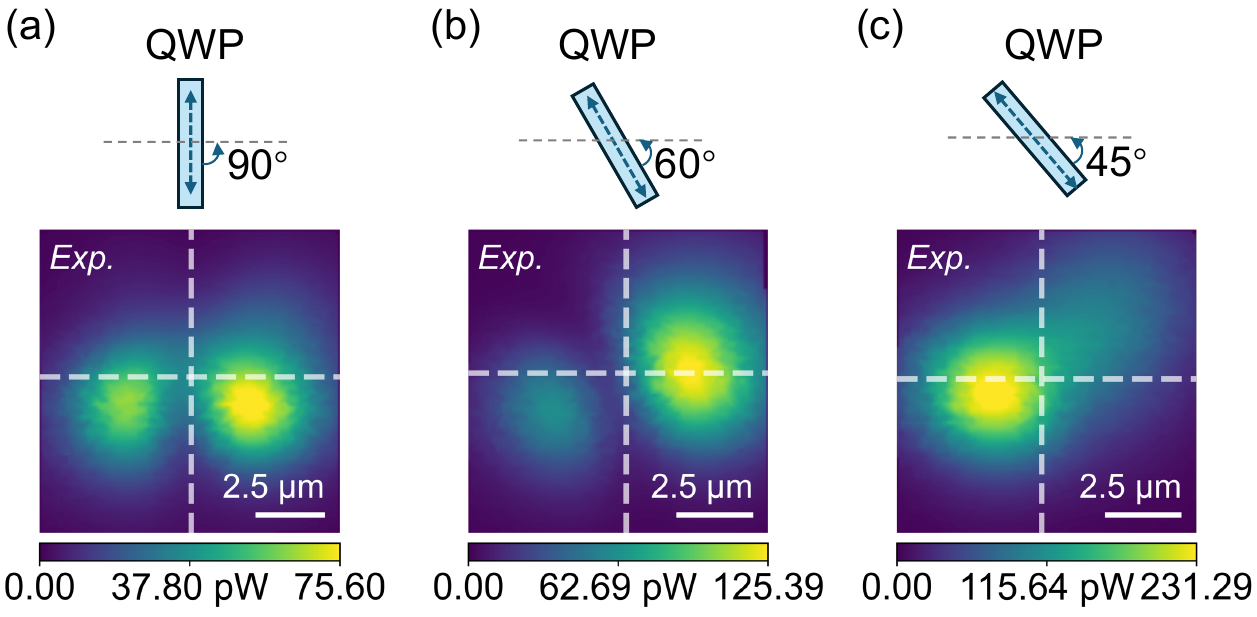}
	\caption{\changes{Measured angle-of-incidence dependent confocal mapping under cross-polarization. The relative angles between the QWP fast axis and $z$-axis are (a) 90$^\circ$, (b) 60$^\circ$ and (c) 45$^\circ$. Here the fast axis is on the $x$-$z$ plane and the initial linear polarization state is set to be parallel to the $x$-axis.}}
	\label{fig:S4}
\end{figure}

\noindent
\changes{By varying an additional degree of freedom-the angle $\theta$ defined by the relative angle of QWP fast axis and the $z$-axis-we observe a similar reorientation of two-lobe pattern. As shown in Fig.~\ref{fig:S4}, the lobe orientation rotates from horizontal at $\theta=90^\circ$ to tilted orientations at $\theta=60^\circ$ and $45^\circ$. This behavior is attributed to an increasing contribution from the HG$_{01}$-like mode as $\cot\theta$ becomes nonzero. 
In addition, we observe an overall increase of the lobe intensity as $\theta$ decreases, revealing a pronounced dependence of the polarization extinction ratio.}

\newpage
\subsection{Direct imaging of the cross-polarized beam pattern}
\begin{figure}[htbp]
	\centering
	\includegraphics[width=0.75\textwidth]{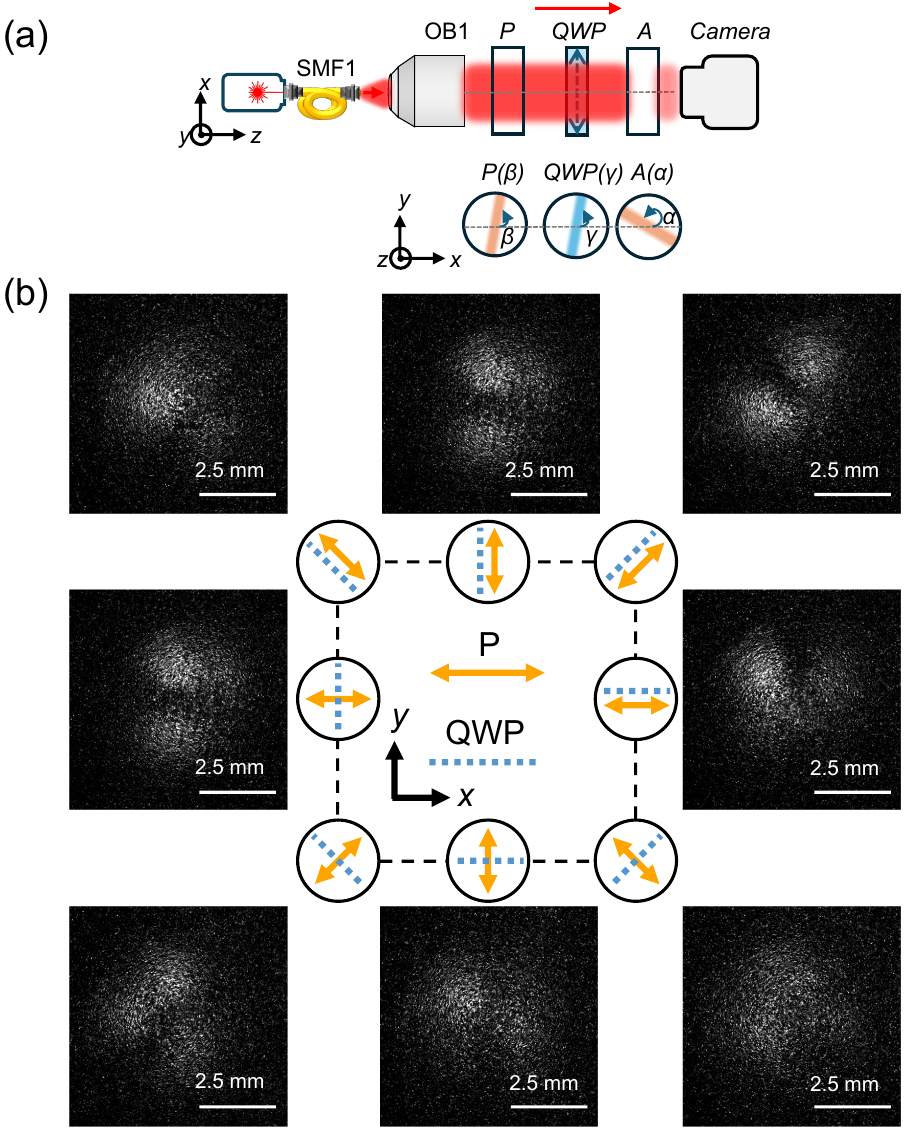}
	\caption{\changes{Direct imaging of the cross-polarized beam pattern. (a) Schematic of the experimental setup used for direct free-space imaging of the beam with a camera. (b) Camera images obtained under 8 different combinations of polarizer and QWP orientations. In all cases, a two-lobe beam structure is clearly observed, confirming that this feature is a signature of spin-orbit interactions by the QWP rather than an artifact of the fiber-based detection scheme.}}
	\label{fig:S5}
\end{figure}

\noindent
\changes{In this section, we show that the observed two-lobe structure is not a detection artifact caused by spatial filtering through the single-mode fiber. We perform a controlled experiment in which the beam profile under cross-polarization is directly recorded by a camera (IDS Imaging, U3-34L0XCP-M-NO) before focusing, as schematically shown in Fig.~\ref{fig:S5}(a). The images in Fig.~\ref{fig:S5}(b) clearly reveal the two-lobe pattern for different polarizer and QWP configurations. This confirms that the observed two-lobe structure is a signature of spin-orbit interactions by the QWP.}

\newpage
\changes{
\subsection{Details for Extracting Fitting Parameters}
\noindent
\paragraph{Residual angular dependence of the QWP retardance.}

In our experiment, the quarter-wave plate (QWP) is nominally operated at normal incidence, which in our notation corresponds to $\theta = 90^\circ$. In this configuration, an ideal wave plate exhibits a purely diagonal Jones matrix: the anti-diagonal contribution $\bar{\bar{Q}}_2$ vanishes as $\cot\theta \to 0$, and the retardance $\Phi_0(\theta)$ reaches an extremum. Consequently, the first angular derivative $\mathrm{d}\Phi_0/\mathrm{d}\theta$ vanishes, and the system is fully described by the conventional Jones matrix of an ideal QWP.

In practice, however, real optical elements deviate from this ideal behavior. Residual first-order angular sensitivity of the retardance can arise from several sources, including the compound zero-order design, small wedge or polishing imperfections, spatial inhomogeneity of birefringence, and a mismatch between the operating and design wavelengths. As a result, $\mathrm{d}\Phi_0/\mathrm{d}\theta$ can remain finite even in the vicinity of normal incidence.

To quantify this effect, we use the angle-dependent retardance data provided by Thorlabs for a quartz zero-order QWP of the same type~\cite{Thorlabs_QWP}, as listed in Table~\ref{tab:QWPRetardance}. From the data points closest to normal incidence ($\pm 0.5^\circ$), we estimate
\[
\left.\frac{\mathrm{d}\Phi_0}{\mathrm{d}\theta}\right|_{\theta \approx 90^\circ}
\approx -0.072,
\]
where both $\Phi_0$ and $\theta$ are expressed in radians. In the following analysis, this value is fixed and treated as a phenomenological parameter, while the remaining model parameters account for the dominant features of the experimentally observed spatial mode transformation.

As a cross-check, we compare this vendor-based estimate with the value obtained from the analytical expression for the phase retardance of a tilted anisotropic plate. Using the standard formula for a quartz plate,
\[
\Phi_0(\theta)=k\left[n_o\,d_o(\theta)-\tilde{n}_e(\theta)\,d_e(\theta)\right],
\]
with
\[
\tilde{n}_e(\theta)=\frac{n_e n_o}{\sqrt{n_e^2 \cos^2 \theta + n_o^2 \sin^2 \theta}},
\]
\[
d_e(\theta)=\frac{\tilde{n}_e(\theta)\,d}{\sqrt{\tilde{n}_e^2(\theta)-\cos^2\theta}},
\qquad
d_o(\theta)=\frac{n_o\,d}{\sqrt{n_o^2-\cos^2\theta}},
\]
where $n_o=1.544$, $n_e=1.553$~\cite{bliokh2016spin}, and $d=23.61~\mu\mathrm{m}$, we obtain a first-order angular slope near normal incidence of approximately
\[
\left.\frac{\mathrm{d}\Phi_0}{\mathrm{d}\theta}\right|_{\theta = 89^\circ} \approx -0.0667.
\]
This value is very close to the estimate extracted from the Thorlabs retardance data, $\mathrm{d}\Phi_0/\mathrm{d}\theta \approx -0.072$. The good agreement between these two independent approaches provides a useful consistency check and supports our choice of fixing $\mathrm{d}\Phi_0/\mathrm{d}\theta$ as a phenomenological input in the fitting procedure.

\paragraph{Fitting procedure.}
The fitting procedure consists of four steps. The input constant parameters are 
$\phi_0=\pi/2$, $\lambda=850$~nm, polarization leakage $b=0.0036$, $\mathrm{d}\Phi_0/\mathrm{d}\theta=-0.072$, and $\theta=90^\circ$.
First, a Gaussian fit of the co-polarized data is performed to determine the beam width $\omega_f$ and the beam center $(x_0, y_0)$. Second, line cuts of the cross-polarized data are taken at $y=y_0$ to obtain initial estimates of $\xi_1$. Third, these initial values are used in the confocal map fitting at maximum extinction to extract $\xi_1$ for each individual dataset. Finally, the individually fitted $\xi_1$ values are used in a global batch fitting of the analyzer-dependent datasets to determine the final value of $\xi_1$.

To quantify the agreement between the measured and fitted intensity distributions, we define two types of relative root-mean-square deviation (RMSD). The peak-normalized RMSD is defined as
\begin{equation}
\mathrm{RMSD}_{\mathrm{peak}} = 
\sqrt{
\frac{
\sum_{i} \left( I^{\mathrm{exp}}_i / I^{\mathrm{exp}}_{\max} 
- I^{\mathrm{fit}}_i / I^{\mathrm{fit}}_{\max} \right)^2
}{
\sum_{i} \left( I^{\mathrm{exp}}_i / I^{\mathrm{exp}}_{\max} \right)^2
}
},
\end{equation}
where $I^{\mathrm{exp}}_i$ and $I^{\mathrm{fit}}_i$ denote the experimental and fitted intensities at pixel $i$, and $I_{\max}$ is the corresponding peak intensity. The energy-normalized RMSD is defined as
\begin{equation}
\mathrm{RMSD}_{\mathrm{energy}} = 
\sqrt{
\frac{
\sum_{i} \left( I^{\mathrm{exp}}_i / \sum_j I^{\mathrm{exp}}_j 
- I^{\mathrm{fit}}_i / \sum_j I^{\mathrm{fit}}_j \right)^2
}{
\sum_{i} \left( I^{\mathrm{exp}}_i / \sum_j I^{\mathrm{exp}}_j \right)^2
}
}.
\end{equation}
The peak-normalized RMSD emphasizes agreement in the spatial profile near the maximum intensity, while the energy-normalized RMSD reflects the global distribution of intensity over the entire field. As shown in Table~\ref{tab:AnalyzerDependent} and Table~\ref{tab:FastAxisDependent}, the peak-normalized RMSD remains low (typically $\sim 0.02$--$0.10$), indicating that the model accurately captures the intensity distribution near the maximum, which is the most critical region for the observed effect. In contrast, the energy-normalized RMSD is consistently higher (typically $\sim 0.08$--$0.32$), suggesting moderate discrepancies in the global intensity distribution, particularly in the low-intensity regions and tails. Overall, the results demonstrate strong agreement in the central spatial features, while the remaining deviations primarily originate from the peripheral intensity distribution.

\begin{table}[htbp]
\centering
\caption{\changes{Angle-dependent retardance of a quartz quarter-wave plate provided by Thorlabs\cite{Thorlabs_QWP}. AOI is given in degrees and retardance in waves. The data are used to estimate the residual first-order angular dependence of the retardance near normal incidence.}}
\label{tab:QWPRetardance}
\setlength{\tabcolsep}{22pt} %
\begin{tabular}{cc|cc}
\toprule
AOI (degree) & Retardance (waves) & AOI (degree) & Retardance (waves)\\
\midrule
-4.5 & 0.2170 & 1.5 & 0.2444 \\
-3.5 & 0.2286 & 2.5 & 0.2374 \\
-2.5 & 0.2359 & 3.5 & 0.2290 \\
-1.5 & 0.2435 & 4.5 & 0.2156 \\
-0.5 & 0.2500 & 5.5 & 0.1974 \\
\phantom{-}0.5 & 0.2498 & 6.5 & 0.1757 \\
\phantom{-}1.0 & 0.2473 &  &  \\
\bottomrule
\end{tabular}
\end{table}

\begin{table}[htbp]
\centering
\setlength{\tabcolsep}{28pt} %
\caption{\changes{Fitting parameters extracted from the analyzer-dependent measurements in Fig.~\ref{fig:2}.
The fitting is performed with $\Phi_0=\pi/2$, $\theta = 90^\circ$ and $\mathrm{d}\Phi_0/\mathrm{d}\theta = -0.072$.}}
\label{tab:AnalyzerDependent}
\begin{tabular}{c c c c}
\toprule
$\delta\alpha$ (degree) & $\xi_1(\mu\mathrm{m})$ & $\mathrm{RMSE}_{\mathrm{peak}}$ & $\mathrm{RMSE}_{\mathrm{energy}}$ \\
\midrule
$-0.16$ & $0.761 - 0.438\,\mathrm{i}$ & $0.040$ & $0.142$ \\
$-0.08$ & $0.761 - 0.438\,\mathrm{i}$ & $0.063$ & $0.228$ \\
$0.00$ & $0.761 - 0.438\,\mathrm{i}$ & $0.058$ & $0.172$ \\
$0.08$ & $0.761 - 0.438\,\mathrm{i}$ & $0.043$ & $0.153$ \\
$0.16$ & $0.761 - 0.438\,\mathrm{i}$ & $0.024$ & $0.087$ \\
\bottomrule
\end{tabular}
\end{table}

\begin{table}[htbp]
\setlength{\tabcolsep}{15pt} %
\caption{\changes{Fitting parameters extracted from the fast axis-dependent measurements in Fig.~\ref{fig:3}.
The fitting is performed with $\Phi_0=\pi/2$, $\theta = 90^\circ$ and $\mathrm{d}\Phi_0/\mathrm{d}\theta = -0.072$.}}
\label{tab:FastAxisDependent}
\centering
\begin{tabular}{c c c c c}
\toprule
Polarizer & Fast Axis of QWP & $\xi_1(\mu\mathrm{m})$ & RMSE$_{\mathrm{peak}}$ & RMSE$_{\mathrm{energy}}$ \\
\midrule
$0^\circ$-Pol & Parallel & $0.761 - 0.438\,\mathrm{i}$ & 0.058 & 0.172 \\
$0^\circ$-Pol & Perpendicular & $-0.665 + 0.054\,\mathrm{i}$ & 0.063 & 0.195 \\
$90^\circ$-Pol & Parallel & $0.344 + 0.970\,\mathrm{i}$ & 0.060 & 0.170 \\
$90^\circ$-Pol & Perpendicular & $-0.575 - 0.394\,\mathrm{i}$ & 0.101 & 0.323 \\
$+45^\circ$-Pol & Parallel & $-0.140 - 0.779\,\mathrm{i}$ & 0.066 & 0.191 \\
$+45^\circ$-Pol & Perpendicular & $-0.032 - 0.819\,\mathrm{i}$ & 0.088 & 0.274 \\
$-45^\circ$-Pol & Parallel & $0.164 - 0.951\,\mathrm{i}$ & 0.061 & 0.168 \\
$-45^\circ$-Pol & Perpendicular & $-0.056 - 0.658\,\mathrm{i}$ & 0.083 & 0.252 \\
\bottomrule
\end{tabular}
\end{table}
}

\begin{figure}[htbp]
	\centering
	\includegraphics[width=0.75\textwidth]{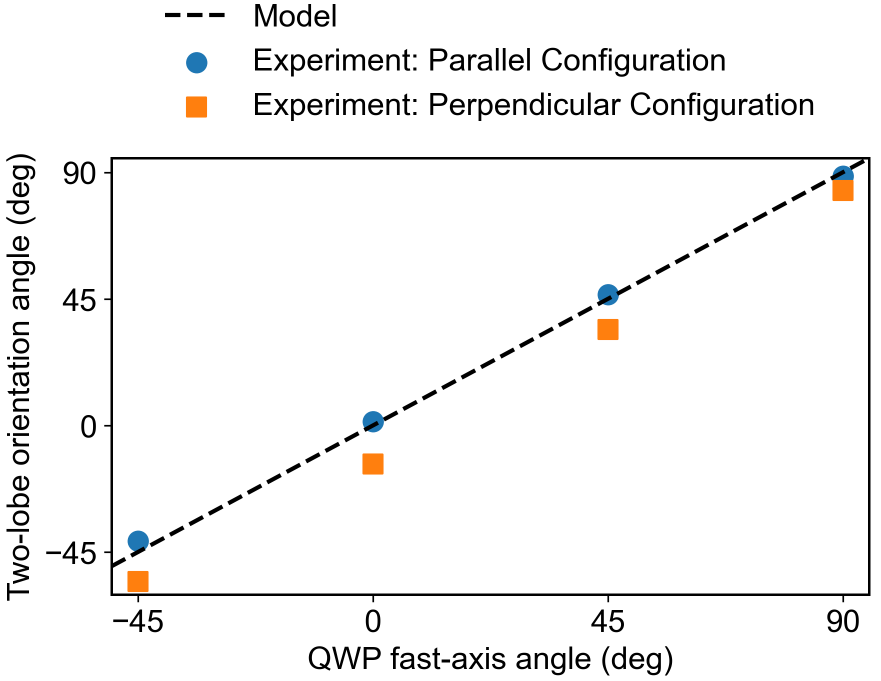}
	\caption{\changes{Comparison between the extracted lobe-axis angle and the QWP fast-axis orientation.The measured lobe orientation is plotted against the QWP fast-axis angle. The dashed line indicates the model prediction. Data are shown for parallel (blue circles) and perpendicular (orange squares) configurations. A linear dependence is observed, with a small offset in the perpendicular case.}}
	\label{fig:S6}
\end{figure}

\clearpage
\changes{
\subsection{Comparison between the extracted lobe-axis angle and the QWP fast-axis orientation}
The extracted orientation of the two-lobe pattern from the experimental confocal maps in Fig.~\ref{fig:3} is plotted against the QWP fast-axis angle in Fig.~\ref{fig:S6}. The dashed line shows the model prediction of a linear dependence. The data are separated into parallel (blue circles) and perpendicular (orange squares) configurations between the polarizer and the QWP fast axis. Both configurations follow the expected trend, although a small offset is observed in the perpendicular case.
}

\bibliography{ref}

\end{document}